\documentclass[numreferences]{kluwer}

\usepackage{epsfig}

\newcommand{\qdot}[1]{\begin{picture}(13,10)
   \put(6,3.6){\circle{13}}
    \put(6,3.6){\makebox(0,0){#1}}
   \end{picture}}
\newcommand{\edot}[1]{\begin{picture}(7,9) 
   \put(3,3.6){\makebox(0,0){#1}}
    \end{picture}}

\begin{document}

\begin{article}
\begin{opening}

\title{Spintronics and Quantum Computing\\ with Quantum Dots}

\author{Patrik Recher and Daniel Loss}

\institute{Department of Physics and Astronomy, University of Basel,
Klingelbergstrasse 82, CH-4056 Basel, Switzerland}

\author{Jeremy Levy}

\institute{Department of Physics
University of Pittsburgh
3941 O'Hara St., Pittsburgh, \\
PA 15260, USA}

\date{}

\begin{abstract}
The creation, coherent manipulation, and measurement of spins in
nanostructures
open up completely new possibilities for electronics and information
processing, among them quantum computing and quantum communication.
We review our theoretical proposal for using electron spins in
quantum dots as quantum bits. We present single- and two qubit gate
mechanisms in laterally as well as vertically coupled quantum dots and
discuss the possibility to couple spins in quantum dots via
superexchange. We further present the recently proposed schemes for
using a single quantum dot as spin-filter and spin read-out/memory device.
\end{abstract}


\end{opening}

\section{Introduction}

Theoretical research on electronic properties in
mesoscopic condensed matter systems has focussed primarily on the
charge degrees of freedom of the electron, while its spin degrees
of freedom have not yet received the same attention. But an
increasing number of spin-related
 experiments\cite{Prinz,Kikkawa,Fiederling,Ohno,Roukes,Ensslin}
show that the
spin of the electron offers unique possibilities for finding novel
mechanisms for information processing and information
transmission---most notably in quantum-confined nanostructures with
unusually long spin dephasing times\cite{Kikkawa,Fiederling,Ohno}
approaching microseconds,
 as well as long distances of up to $100\:\mu{\rm m}$ \cite{Kikkawa}
 over which spins can be transported phase-coherently.
Besides the intrinsic interest in spin-related phenomena, there are
two main areas which hold promises for future applications:
Spin-based devices in conventional\cite{Prinz} as well as in quantum
computer hardware\cite{Loss97}.
In conventional computers, the
electron spin can be expected to enhance the performance of quantum
electronic devices, examples being spin-transistors (based on
spin-currents and spin injection), non-volatile memories,
single spin as the ultimate limit of information storage
etc.\cite{Prinz}.
On the one hand, none
of these devices exist yet, and experimental progress as well as
theoretical investigations are needed to provide guidance and
support in the search for realizable implementations. On the other
hand, the emerging field of quantum computing\cite{Steane,MMM2000} and
quantum
communication\cite{MMM2000,Bennett00} requires a radically new
approach to the design of the necessary hardware. As first pointed out
in Ref.\cite{Loss97},
 the spin of the electron is a most natural candidate for the
qubit---the
fundamental unit of quantum information. We have shown\cite{Loss97} that
these spin qubits, when located in quantum-confined structures such
as semiconductor quantum dots, atoms or molecules, satisfy all
requirements needed for a scalable quantum computer. Moreover, such
spin-qubits---being attached to an electron with orbital degrees of
freedom---can be transported along conducting wires between
different subunits in a quantum network\cite{MMM2000}. In particular,
spin-entangled electrons can be created in coupled quantum dots and---as
mobile Einstein-Podolsky-Rosen (EPR) pairs\cite{MMM2000}---provide then
the necessary resources for quantum communication.

It follows a short introduction of quantum computing  and quantum
communication and we will then present our current theoretical efforts
towards a realization of quantum computing. We thereby focus on the
implementation of the necessary gate and read-out operations schemes
with quantum dots.

\subsection{Quantum Computing and Quantum Communication}
\label{ssecQC}
We give a brief description of the emerging research field of quantum
computation.
It has attracted much interest
 recently
 as it opens up the possibility of outperforming classical
 computation through new and more powerful quantum algorithms such
 as the ones discovered by Shor\cite{Shor94} and by Grover\cite{Grover}.
There is now
 a growing list of quantum tasks\cite{MMM2000,Bennett00} such as
cryptography, error
 correcting schemes,  quantum teleportation, etc. that have
 indicated even more the desirability of experimental
 implementations of quantum computing.
In a quantum computer each quantum bit (qubit) is allowed to be
 in any state of a quantum two-level system.
All quantum algorithms can be implemented by concatenating one-
 and two-qubit gates.
 There is a growing number of proposed physical
 implementations of qubits and quantum gates. A few examples are:
 Trapped ions\cite{traps},
 cavity QED\cite{cavity},
 nuclear spins\cite{nmr,Kane},
 superconducting devices\cite{Schon,Averin,Ioffe,Mooij},
 and our qubit proposal\cite{Loss97}  based on the spin of the electron
 in quantum-confined nanostructures.

\subsection{Quantum Dots}
\label{ssecQD}
Since quantum dots are the central objects of this work we shall make
some general remarks about these systems here.
Semiconductor quantum dots are structures where charge carriers
 are confined in all three spatial dimensions,
 the dot size being of the order of the Fermi wavelength
 in the host material,
 typically between $10\:{\rm nm}$ and $1\:{\rm \mu m}$
\cite{kouwenhoven}.
The confinement is usually achieved by electrical gating of a
 two-dimensional electron gas (2DEG),
 possibly combined with etching techniques.
Precise control of the number of electrons in the conduction band
 of a quantum dot (starting from zero) has been achieved in GaAs
 heterostructures\cite{tarucha}.
The electronic spectrum of typical quantum dots can vary strongly
 when an external magnetic field is applied\cite{kouwenhoven,tarucha},
 since the magnetic length corresponding to typical laboratory fields
 $B\approx 1\,{\rm T}$ is comparable to typical dot sizes.
In coupled quantum dots
 Coulomb blockade effects\cite{waugh},
 tunneling between neighboring dots\cite{kouwenhoven,waugh},
 and magnetization\cite{oosterkamp} have been observed as well as the
 formation of a delocalized single-particle state\cite{blick}.

\section{Quantum Gate Operations with Coupled Quantum Dots}
\label{coupling}

One and two qubit gates are known to be sufficient to carry out any
quantum algorithm. For electron spins in nearby coupled quantum dots the
desired two qubit coupling is provided by a combination of Coulomb
interaction and the Pauli exclusion principle.

At zero magnetic field, the ground state of two
coupled electrons is a spin singlet,
whereas the first excited state in the presence of strong Coulomb
repulsion is usually a triplet. The remaining spectrum is separated
from these two states by a gap which is either given by the Coulomb
repulsion or the single particle confinement.
The low-energy physics of such a system can then be described by
the Heisenberg spin Hamiltonian
\begin{equation}\label{Heisenberg}
H_{\rm s}(t)=J(t)\,\,{\bf S}_1\cdot{\bf S}_2,
\end{equation}
where $J(t)$ is the exchange coupling  between
the two spins ${\bf S}_{1}$ and ${\bf S}_{2}$, and is given by the
energy
difference between the singlet and triplet states.
If we pulse the exchange coupling such that $\int dtJ(t)/\hbar =
J_0\tau_s/\hbar = \pi$ (mod $2\pi$), the associated unitary time
evolution $U(t) = T\exp(i\int_0^t H_{\rm s}(\tau)d\tau/\hbar)$
corresponds to the ``swap'' operator $U_{\rm sw}$ which
exchanges the quantum states of qubit 1 and 2 \cite{Loss97}. Having an
array of dots it is therefore possible to couple any two qubits.
Furthermore,
the quantum XOR gate can be constructed by applying the
sequence\cite{Loss97}
\begin{equation}
U_{\rm XOR} = e^{i(\pi/2)S_1^z}e^{-i(\pi/2)S_2^z}U_{\rm sw}^{1/2}e^{i\pi
S_1^z}U_{\rm sw}^{1/2},
\end{equation}
i.e. a combination of
``square-root of swap'' $U_{\rm sw}^{1/2}$ and single-qubit rotations
$\exp(i\pi S_i^z)$. Since $U_{\rm XOR}$ (combined with
single-qubit rotations) is proven to be a universal quantum
gate\cite{Barenco}, it can be used to assemble any quantum algorithm.
The study of universal quantum computation in coupled quantum dots
is thus essentially reduced to the study of single qubit rotations and
the {\it exchange mechanism}, in particular how the exchange coupling
$J(t)$ can be controlled experimentally.
Note that the switchable coupling mechanism described
below need not be restricted to quantum dots: the same
principle can be used in other systems, e.g. coupled atoms in a
Bravais lattice, supramolecular structures, or
overlapping shallow donors in semiconductors.

\subsection{Laterally coupled quantum dots}

We first discuss a system of two laterally coupled quantum dots
defined by depleted regions in a 2DEG containing one (excess) electron
each\cite{Burkard}. The electrons are allowed to tunnel between the dots
(if the tunnel barrier is low) leading to spin correlations  via their
charge (orbital) degrees  of freedom.
We model the coupled system with the Hamiltonian
$H = H_{\rm orb} + H_{\rm Z}$, where $H_{\rm orb} = \sum_{i=1,2} h_i+C$
with
\begin{eqnarray}
h_i = \frac{1}{2m}\left({\bf p}_i-\frac{e}{c}{\bf A}({\bf r}_i)
\right)^2+V({\bf r}_i),\,\,\,\,\,
C={{e^2}\over{\kappa\left| {\bf r}_1-{\bf r}_2\right|}}\,\,\,\,.
\label{hamiltonian}
\end{eqnarray}
Here, $h_i$ describes the single-electron dynamics in the 2DEG
confined to the $xy$-plane, with $m$ being the effective electron mass.
We allow for a  magnetic field ${\bf B}= (0,0,B)$ applied along the
$z$-axis that couples to the electron charge via the
vector potential ${\bf A}({\bf r}) = \frac{B}{2}(-y,x,0)$, and to the
spin
via a Zeeman coupling term $H_{\rm Z}$.
The single dot confinement as well as the tunnel-coupling is modeled by
a quartic
potential,
$V(x,y)=\frac{m\omega_0^2}{2}\left(\frac{1}{4 a^2}\left(x^2-a^2
\right)^2+y^2\right)$,
which, in the limit $a\gg a_{\rm B}$,  separates  into two harmonic
wells  of
frequency $\omega_0$ where $2a$ is the interdot distance and
$a_{\rm B}=\sqrt{\hbar/m\omega_0}$
is the effective Bohr radius of a dot.
This choice for the potential is motivated by the experimental
observation\cite{tarucha} that
the low-energy spectrum of single dots is well described by a parabolic
confinement potential.
The (bare) Coulomb interaction between the two electrons is
described by $C$ where $\kappa$ denotes the dielectric constant of the
semiconductor. The screening length $\lambda$ in almost depleted regions
like few-electron quantum dots can be expected to be much larger than
the
bulk 2DEG screening length (about $40\,{\rm nm}$ for GaAs).
Therefore, $\lambda$ is large compared to the size of the coupled
system,
$\lambda\gg 2a\approx 40\,{\rm nm}$
for small dots, and we will consider the limit of unscreened
Coulomb interaction
($\lambda/a\gg 1$).
At low temperatures $kT_{B}\ll \hbar\omega_0$ we are allowed
to restrict our analysis to the two lowest orbital eigenstates of
$H_{\rm orb}$, leaving us with a symmetric (spin-singlet) and an
antisymmetric
(three triplets $T_{0}$, $T_{\pm}$) orbital state.
In this reduced (four-dimensional) Hilbert space, $H_{\rm orb}$
can be replaced by the effective Heisenberg spin Hamiltonian
Eq.~(\ref{Heisenberg}) where
the exchange coupling $J=\epsilon_{\rm t}-\epsilon_{\rm s}$ is
given by the difference between the triplet and
singlet energy. We make use of the analogy between atoms
and quantum dots (artificial atoms) and caculate
$\epsilon_{\rm t}$ and $\epsilon_{\rm s}$ with variational methods
similiar to the ones used in molecular physics.
With the Heitler-London approximation using single-dot groundstate
orbitals we find\cite{Burkard},
\begin{eqnarray}\label{J}
J &=& \frac{\hbar\omega_0}{\sinh\left(2d^2\,\frac{2b-1}{b}\right)}
\Bigg\{
\frac{3}{4b}\left(1+bd^2\right)\\ \nonumber
 && + c\sqrt{b} \left[e^{-bd^2} \, I_0\left(bd^2\right)
- e^{d^2 (b-1)/b}\, I_0\left(d^2\,\frac{b-1}{b}\right)\right]
\Bigg\},
\end{eqnarray}
where we introduce the dimensionless distance $d=a/a_{\rm B}$ and
the magnetic compression factor
$b=B/B_0=\sqrt{1+\omega_L^2/\omega_0^2}$,
where $\omega_L=eB/2mc$ is the Larmor frequency.
\begin{figure}
\centerline{\psfig{file=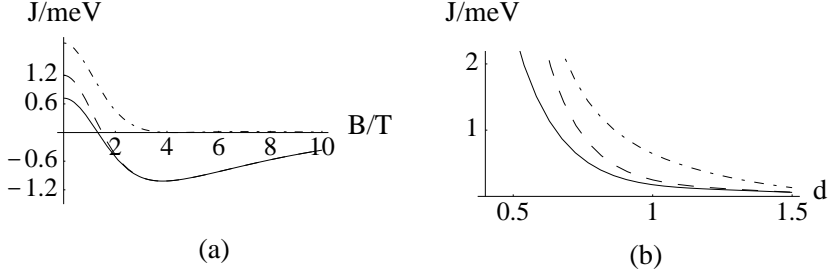,width=11cm}}
\caption[]{
The exchange coupling $J$ (full line) for GaAs quantum dots with
confinement energy $\hbar\omega=3\,{\rm meV}$ and $c=2.42$.
For comparison we plot the usual short-range Hubbard result $J=4t^2/U$
(dashed-dotted line) and the extended Hubbard result\cite{Burkard}
$J=4t^2/U+V$. In (a), $J$ is plotted as a
function of magnetic field $B$ at fixed inter-dot distance
($d=a/a_{\rm B}=0.7$), and in (b) as a function of the
inter-dot distance $d=a/a_{\rm B}$ at $B=0$.}
\label{Jplots}
\end{figure}
${\rm I_0}$ denotes the zeroth Bessel function.
The first term in Eq.~(\ref{J}) comes from the confinement potential.
The terms
proportional to $c=\sqrt{\pi/2}(e^2/\kappa a_{\rm B})/\hbar\omega_0$ are
due to the Coulomb interaction $C$, where the exchange term appears with
a minus sign.
Note that typically $|J/\hbar\omega_0|\ll 1$ which makes the exclusive
use of ground-state single-dot orbitals in the Heitler-London ansatz
a self-consistent approach.
The exchange $J$ is given as a function of $B$ and $d$ in
Fig.~\ref{Jplots}.
We observe that $J>0$ for $B=0$, which is generally true for a
two-particle system with time reversal invariance.
The most remarkable feature of $J(B)$, however, is the change of sign
from positive (ferromagnetic) to negative (antiferromagnetic), which
occurs at some finite $B$ over a wide
range of parameters $c$ and $a$.
This singlet-triplet crossing is caused by the long-range Coulomb
interaction and is therefore absent in the standard Hubbard model that
takes only into account short range interaction and, in the limit
$t/U\ll 1$,  is given by $J=4t^2/U>0$ (see Fig.~\ref{Jplots}).
Large magnetic fields ($b\gg 1$) and/or large interdot distances ($d\gg
1$) reduce the overlap between the dot orbitals leading to an
exponential decay of $J$ contained in the $1/\sinh$ prefactor in
Eq.~(\ref{J}).
This exponential suppression is partly compensated
by the exponentially growing exchange term $\propto
\exp(2d^2(b-1/b))$. As a consequence, $J$ decays
exponentially as $\exp(-2d^2b)$ for large $b$ or $d$.
Thus, $J$ can be tuned through zero and then exponentially suppressed to
zero by a magnetic field in a
very efficient way (exponential switching is highly desirable to
minimize
gate errors). Further, working around the singlet-triplet crossing
provides a smooth exchange switching, requiring only small local
magnetic fields.
Qualitatively similar results are obtained\cite{Burkard} when we extend
the Heitler-London result by taking into account higher levels and double
occupancy of the dots (using a Hund-Mullikan approach).
In the absence of tunneling ($J=0$) direct Coulomb interaction between
the electron charges can still be present. However the spins (qubit)
remain unaffected provided the spin-orbit coupling is sufficiently
small, which is the case for s-wave electrons in GaAs structures with
unbroken inversion symmetry.
Finally, we note that a spin coupling can also be achieved on a long
distance scale by using a cavity-QED scheme\cite{Imamoglu} or
superconducting leads to which the quantum dots are attached\cite{CBL}.

\subsection{Vertically coupled quantum dots}

We also investigated vertically coupled Quantum dots\cite{vertical}.
This kind of coupling can be implemented with multilayer self-assembled
quantum dots\cite{luyken} as well as with etched mesa
heterostructures\cite{austing}.

\begin{figure}[t]
\begin{tabular}{l r}
\psfig{file=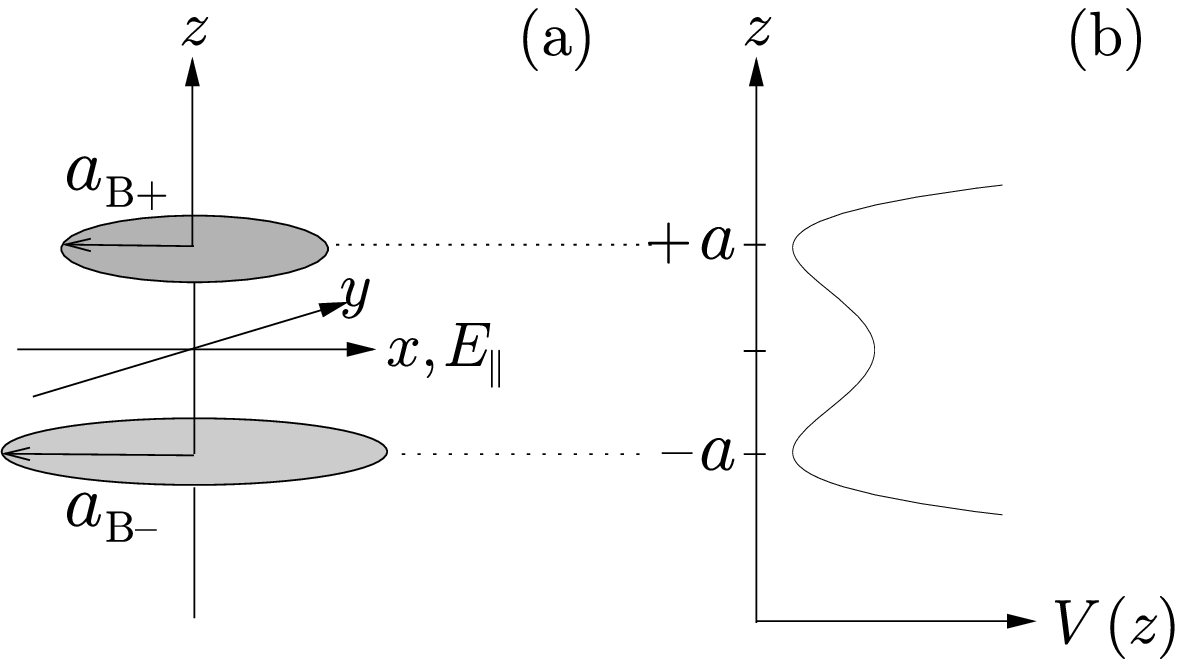,width=5.5cm} &
\psfig{file=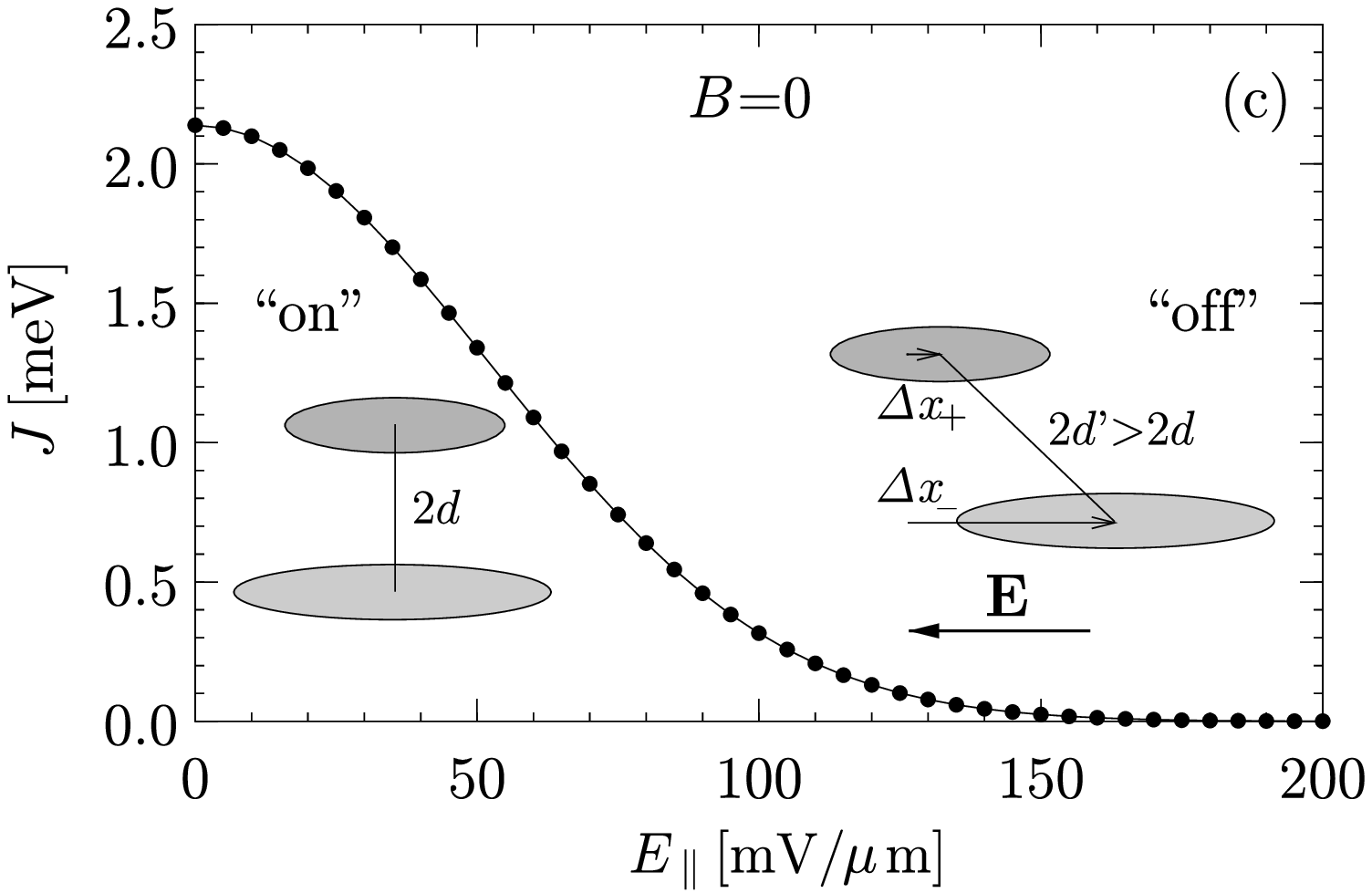,width=5.5cm}
\end{tabular}
\caption[]{
(a) Sketch of the vertically coupled double quantum-dot
system. The two dots may have different lateral diameters,
$a_{B+}$ and $a_{B-}$. We consider an in-plane electric field $E_\parallel$.
(b) The model potential for the vertical confinement
is a double well, which is obtained by combining two
harmonic wells with frequency $\omega_z$ at $z=\pm a$.
(c) Switching of the spin-spin coupling between dots of different
size by means of an in-plane electric field $E_\parallel$ ($B=0$).
The exchange coupling is switched ``on'' at $E=0$ (see text).
We have chosen $\hbar\omega_z=7\,{\rm meV}$, $d=1$,
$\alpha_{0+}=1/2$ and $\alpha_{0-}=1/4$.
For these parameters,
$E_0=\hbar\omega_z/ea_B=0.56\,{\rm mV/nm}$ and
$A=(\alpha_{0+}^2-\alpha_{0-}^2)/2\alpha_{0+}^2\alpha_{0-}^2=6$.
The exchange coupling $J$ decreases exponentially on the scale
$E_0/2A = 47 \,{\rm mV/\mu m}$ for the electric field.
}
\label{vdots}
\end{figure}

We model the vertical coupled dot system by a potential $V=V_l+V_v$
where $V_{l}$ describes the parabolic lateral confinement and $V_{v}$
models the vertical dot coupling assumed to be a quartic potential
similar to the one introduced above for the lateral coupling. We allow
for different dot sizes $a_{{\rm
B}\pm}=\sqrt{\hbar/m\alpha_{0\pm}\omega_z}$ with $\omega_{z}$ being the
vertical confinement (see Fig. \ref{vdots}), implying an effective Bohr
radius $a_{\rm B}=\sqrt{\hbar/m\omega_z}$ and
 a dimensionless interdot distance $2d = 2a/a_{\rm B}$. By applying an
in-plane electric field $E_\parallel$ (see Fig. \ref{vdots}) an
interesting new switching mechanism arises. The dots are shifted
parallel to the field
  by $\Delta x_\pm =E_\parallel/E_0\alpha_{0\pm}^2$, where
$E_0=\hbar\omega_z/ea_B$.
Thus, the larger dot is shifted a greater distance $\Delta x_{-}>\Delta
x_{+}$
 and so the mean distance between the electrons grows as
 $d'=\sqrt{d^2+A^2(E_\parallel/E_0)^2}>d$, taking
 $A=(\alpha_{0+}^2-\alpha_{0-}^2)/2\alpha_{0+}^2\alpha_{0-}^2$.
Since the exchange coupling $J$ is exponentially sensitive to the
interdot distance $d'$ (see Eq. (\ref{J}))
we have another exponential switching
mechanism for quantum gate operations at hand.

\subsection{coupling two spins by superexchange}

There is a principal problem if one wants to couple two ``extended"
dots  whose energy
levels are closely spaced (i.e. smaller than $k_BT$), as would 
be the case
if there is a sizable distance between the
two
confined qubits before the barrier is lowered.
 In this case, the
singlet-triplet splitting
becomes
vanishingly small, and it would not take much excitation energy to get
states which are not
entangled at all. In other words, the adiabatic switching
time\cite{Burkard} which is proportional to the
inverse
level spacing becomes arbitrarily large. A better scenario for coupling
the two
spin-qubits is to
make use of a superexchange mechanism to obtain a Heisenberg
interaction\cite{Loss97}. Consider three aligned quantum
dots where the middle dot is empty and 
so small that only its lowest
levels  will be occupied by
1 or 2 electrons in a virtual
hopping process (see Fig. 3).
The left
and right dots can be much larger
but still small enough such that the
Coulomb charging energies $U_{L}\approx U_{R}$  are high enough
(compared to $k_BT$) to suppress any double occupancy. Let
us assume now that the middle
dot has  energy levels higher than the 
ground states of right
and left
dots, assumed to be approximately the same. These levels
include single particle energy (set to zero) and
Coulomb charging energy $N^2e^2/2C$, with $N$ the number of electrons
and C the capacitance of the middle dot,
and thus  the ground state energy of the middle dot is $0$ when
empty, $\epsilon=e^2/2C$  for one electron, and $4\epsilon$ for 2 electrons.
The tunnel coupling between the dots is denoted by $t_{0}$. Now,
there
are two types of virtual processes possible which couple the spins but
only one is dominant.
First,
the electron of the left (right) dot hops on the middle dot, and then
the electron from the
right
(left) dot hops on the {\it same} level on the middle dot, and thus, due
to the Pauli principle, the
two electrons on the middle dot form a singlet, giving the desired
entanglement. And then they
hop
off again into  the left and right dots, respectively. (Note that U must be
larger than $k_{B}T$,
otherwise
real processes involving 2 electrons in the left or right dot will be
allowed). It is not
difficult to
see that this virtual process leads to an effective Heisenberg exchange
interaction with exchange constant $J=4t_{0}^4/4\epsilon^3$, where the
virtual energy denominators follow the sequence $1/\epsilon\rightarrow
1/4\epsilon\rightarrow
1/\epsilon$. 

In the second type of virtual process the left (right)
electron hops via the middle dot into
the right (left) dot and forms there a singlet, giving
$J=4t_{0}^4/U_{R}\epsilon^2$.
\begin{figure}
\label{levy}
\centerline{\psfig{figure=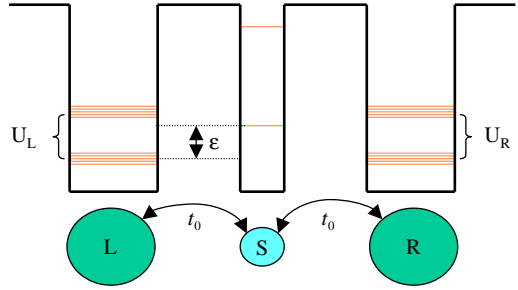,width=7cm}}
\caption{Geometry for superexchange method of coupling two quantum
dots.}
\end{figure}
However, this process has
vanishing weight because there are also many nearby states available in
the outer dots
for which
there is no spin correlation required by the Pauli principle. Thus, most
of the virtual
processes,
for which we have 2 electrons in the left (right) dot, do not produce spin
correlations, and
thus we
can neglect these virtual processes of the second type altogether.
It should be possible to create ferroelectrically defined nanostructures
for which
superexchange is the dominant mechanism for coupling neighboring
electrons. The
geometry
will resemble closely that of Fig. 3, except that the central barrier
becomes a
narrow well.

\section{Single-Spin Rotations}

In order to perform one qubit gates single-spin rotations are required.
This is done by exposing  a single spin to a time-varying Zeeman
coupling
 $(g\mu_B {\bf S}\cdot {\bf B})(t)$ \cite{Burkard},
 which can be controlled through
 both the magnetic field ${\bf B}$ and/or the g-factor $g$.
We have proposed a number of possible
implementations\cite{Loss97,Burkard,MMM2000,BEL} for spin-rotations:
Since only relative phases between qubits are relevant we can apply  a
homogeneous ${\bf B}$-field rotating all spins at once. A local change
of the Zeeman coupling is then possible by changing the Larmor frequency
$\omega_{L}=g\mu_{B}B/\hbar$. The equilibrium position of an electron
can be changed through electrical gating, therefore if the electron
wavefunction is pushed into a region with different magnetic field
strength or different (effective) g-factor, the relative phase of such
an electron then becomes $\phi = (g'B'-gB)\mu_B\tau/2\hbar$. Regions
with an increased magnetic field can be provided
 by a magnetic (dot) material  while an effective magnetic field can be
produced e.g.\ with dynamically polarized nuclear spins (Overhauser
effect)\cite{Burkard}.
We shall now explain a concept for using g-factor-modulated
 materials\cite{MMM2000,BEL}.
In bulk semiconductors
 the free-electron value of the Land\'e g-factor $g_0=2.0023$
 is modified by spin-orbit coupling. Similarly, the g-factor
can be drastically enhanced by doping the semiconductor
with magnetic impurities\cite{Ohno,Fiederling}.
In confined structures such as quantum wells, wires, and dots,
 the g-factor is further modified and becomes sensitive
 to an external bias voltage\cite{Ivchenko}.
We have numerically analyzed a system with a layered structure
 (AlGaAs-GaAs-InAlGaAs-AlGaAs),
 in which the effective g-factor of electrons is varied by
 shifting their equilibrium position from one layer to
 another by electrical gating\cite{Ensslin2}.
We have found that in this structure the effective g-factor
 can be changed by about $\Delta g_{\rm eff}\approx 1$~\cite{BEL}.

Alternatively one can use electron-spin-resonance (ESR) techniques\\
\cite{Burkard}  to perform single-spin rotations, e.g. if we want to
flip a certain qubit (say from $|\uparrow\rangle$ to
$|\downarrow\rangle$) we apply an ac-magnetic field perpendicular to the
$\uparrow$- axis that matches the Larmor frequency of that particular
electron.  Due to paramagnetic resonance\cite{Shankar} the spin can
flip.

 Furthermore, localized magnetic fields can be generated
with the magnetic tip of a scanning force microscope,
a magnetic disk writing head,
by placing the dots above a grid of current-carrying wires,
 or by placing a small wire coil above the dot etc.

\section{Read-Out of a Single Spin}
The final step of each (quantum) computation, consists in reading out
the state of each qubit, i.e. if the electron spin is in the
$|\uparrow\rangle$ or $|\downarrow\rangle$ state. It is very hard to
detect an electron spin over its tiny (of the order of $\mu_{B}$)
magnetic moment directly.  We proposed several devices for read out like
tunneling of the electron into a supercooled paramagnetic
dot\cite{Loss97,MMM2000}, thereby inducing a magnetization nucleation
from the metastable phase into a ferromagnetic domain. The domain's
magnetization direction is along the measured spin and can be detected
by conventional methods and provides a 75\%-reliable result for the read
out of the electron spin. Another possibility is to use a spin selective
tunnelbarrier (conventional spin filter), that let pass only one spin
direction. If an electron passes the barrier to enter another dot an
electrometer can detect the charge\cite{Loss97}.

\subsection{Quantum Dot As Spin Filter and\\ \qquad\,Read-Out/Memory
Device}

We recently proposed\cite{Recher} another setup using a quantum dot
attached to in and outgoing current leads $l=1,2$---that can work either
as a spin filter or as a read-out device, or as a spin memory where the
spin stores the information.
A new feature of this proposal is that we lift the spin-degeneracy
 with {\it different} Zeeman splittings for the dot and
 the leads,
 e.g.\ by using materials with different effective
g-factors for leads
and dot\cite{Recher}.
This results in Coulomb
 blockade oscillation peaks and spin-polarized currents which are
uniquely associated
with the spin state
 on the dot.

The setup is described by a standard tunneling  Hamiltonian
$H_0+H_T$ \cite{Mahan},
 where $H_0=H_L+H_D$ describes the leads and the dot.
$H_D$ includes
 the charging and interaction energies of the electrons in the dot
 as well as their Zeeman energy $\pm g\mu_B B/2$
 in an external magnetic field ${\bf B}$.
Tunneling between leads and the dot is described by
$H_T=\sum_{l,k,p,\sigma}t_{lp}c_{lk\sigma}^{\dag}d_{p\sigma}+{\rm
h.c.}$,
 where $c_{lk\sigma}$ annihilates electrons with spin $\sigma$ and
momentum $k$ in lead~$l$
 and $d_{p\sigma}$ annihilates electrons in the dot. We work in
the Coulomb blockade regime\cite{kouwenhoven} where the charge on the
dot
is quantized. We use a stationary master equation
approach\cite{kouwenhoven,Recher}  for the reduced density matrix of the
dot and calculate the
 transition rates in
 a ``golden-rule'' approach up to 2nd order in $H_T$.
The first-order contribution to the current is
 the sequential tunneling (ST) current $I_s$\cite{kouwenhoven},
 where the number of electrons on the dot fluctuates
 and thus the processes of an electron tunneling from the lead onto the
dot
 and vice versa are allowed by energy conservation.
The second-order contribution is
 the cotunneling (CT) current $I_c$\cite{averinnazarov},
 where  charge is transported over intermediate virtual states of the
dot.

We now consider a system,
where the Zeeman splitting in the leads is negligible (i.e.\
much smaller than the Fermi energy)
while on the dot it is given as $\Delta_z = \mu_B |gB|$.
We assume  a small bias $\Delta\mu = \mu_1-\mu_2 >0$
 between the leads at chemical potential $\mu_{1,\,2}$
 and low temperatures so that $\Delta\mu,\, k_{B}T < \delta$,
 where $\delta$ is the characteristic energy-level distance on the dot.
First we tune the system to the ST resonance $\mu_{1}>\Delta E>\mu_{2}$
where the number of electrons can fluctuate between $N$  and $N+1$.
$\Delta E$ is the energy difference between the $N+1$ and $N$-particle
groundstates (GS) of the dot. We first consider a quantum dot with $N$
odd and total spin $s=1/2$ with the $N$-particle GS to be
$|\uparrow\rangle$ and to have energy $E_{\uparrow}=0$. In this state
the dot can receive an electron from the leads and, depending on the
spin of the incoming electron form a singlet $|S\rangle$ with energy
$E_{S}$ (for spin down) or a triplet $|T_{+}\rangle$ with energy
$E_{T_{+}}$ (for spin up). The singlet is (usually) the GS for $N$ even,
whereas the three triplets $|T_{\pm}\rangle$ and $|T_{0}\rangle$ are
excited states. In the regime
$E_{T_{+}}-E_{S},\,\Delta_{z}>\Delta\mu,\,k_{B}T$,  energy conservation
only allows ground state transitions.
Thus, spin-up electrons are not allowed to tunnel from lead~$1$
 via the dot into lead~$2$,
 since this would involve virtual states $|T_{+}\rangle$ and
$|\downarrow\rangle$,
 and so we have $I_s(\uparrow)=0$ for ST.
However, spin down electrons may pass through the dot in
 the process
 \edot{$\downarrow$}\qdot{$\uparrow$}$_i$~$\to$
 \qdot{$\uparrow\downarrow$}$_f$,
followed by
 \qdot{$\uparrow\downarrow$}$_i$~$\to$
 \qdot{$\uparrow$}\edot{$\downarrow$}$\!_f$.
Here the state of the quantum dot is drawn inside the circle,
 while the states in the leads are drawn to the left and right, {\it
resp.},
 of the circle.
This leads to a {\it spin-polarized}
 ST current $I_s = I_s(\downarrow)$,
 which we have calculated as\cite{Recher}
\begin{eqnarray}
&& I_s(\downarrow)/I_0=\theta(\mu_1-E_S)-\theta(\mu_2-E_S), \quad
k_B T<\Delta\mu ,
\label{eqnSmallT} \\
&& I_s(\downarrow)/I_0=
\frac{\Delta\mu}{4k_BT}\cosh^{-2}\left[\frac{E_S-\mu}{2k_BT}\right],
\quad k_BT >\Delta\mu,
\label{eqnLargeT}
\end{eqnarray}
where $\mu = (\mu_1+\mu_2)/2$
 and $I_0=e\gamma_1\gamma_2/(\gamma_1+\gamma_2)$.
Here $\gamma_l=2\pi\nu|A_{lnn'}|^2$ is the tunneling rate
 between lead~$l$ and the dot. $n$ and $n'$ denote the $N$ and $N+1$
particle eigenstates  of $H_{D}$ involved in the tunnel process. The
dependence of
$A_{ln'n}=\sum_{p\sigma}t_{lp}\langle  n'|d_{p\sigma}| n\rangle$ on $n$
and $n'$ is weak compared to the resonant character of the tunneling
current considered here\cite{Recher}.
Similarly, for $N$ even we find $I_s(\downarrow)=0$
while for $I_s(\uparrow)$ a similar result holds\cite{Recher} as
 in Eqs. (\ref{eqnSmallT}), (\ref{eqnLargeT}).

Even though $I_s$ is completely spin-polarized,
 a leakage of current with opposite polarization
 arises through cotunneling processes\cite{Recher};
still the leakage is small, and the efficiency
for $\Delta_z<|E_{T_+}-E_S|$ for spin filtering in
the sequential regime becomes\cite{Recher}
\begin{equation}
\label{efficiencyST}
I_s(\downarrow)/I_c(\uparrow)\sim
\frac{\Delta_z^2}{(\gamma_1+\gamma_2)\max\{k_BT,\Delta\mu\}},
\end{equation}
 and equivalently for
 $I_s(\uparrow)/I_c(\downarrow)$ at the even-to-odd transition.
In the ST regime
 we have $\gamma_i< k_{B}T,\Delta\mu$,
 thus, for $k_{B}T,\Delta\mu<\Delta_z$,
 we see that the spin-filtering is very efficient.
Above or below a ST-resonance the system is in the CT regime where the
current is solely due to CT-processes. Again, in the regime
$E_{T_{+}}-E_{S},\,\Delta_{z}>\Delta\mu,\,k_{B}T$ the current is   {\it
spin-polarized} and the spin filter also works in the CT
regime\cite{Recher}.

We discuss now the opposite case where the leads are fully spin
polarized
 with a much smaller Zeeman splitting on the dot\cite{Recher}.
Such a situation can be realized with magnetic
semiconductors (with effective g-factors reaching
100 \cite{Fiederling}) where spin-injection into GaAs has recently been
demonstrated for the first time\cite{Fiederling,Ohno}.
Another possibility would be to work in the quantum Hall regime
 where spin-polarized edge states are coupled to a quantum
 dot\cite{Sachrajda}.
In this setup the device can be used as read-out for the spin state
on the dot.
Assume now that the spin polarization in both leads is up,
and the ground state of the dot contains an odd
 number of electrons with total spin $1/2$.
Now the leads can {\it provide} and {\it take up} only
spin-up electrons. As a consequence,  a ST
current will only be possible if the dot state is $|\downarrow\rangle$
(to form a
singlet with the incoming electron, whereas the triplet is excluded by
energy conservation). Hence,
the current is much larger for the spin on the dot being in
$|\downarrow\rangle$
 than it is for $|\uparrow\rangle$. Again, there is a small CT leakage
current for the dot-state $|\uparrow\rangle$, with a ratio of the two
currents given by Eq. (\ref{efficiencyST}) (assuming
$E_{S}>\Delta_{z}$).
Thus, we can probe (read out) the
  spin-state on the quantum dot by measuring the current
which passes through the dot. Given that the
ST current is typically on the order of  $0.1-1$
nA \cite{kouwenhoven}, we can
estimate the read-out frequency $I/2\pi e$ to be on the order of $0.1-1$
GHz.
Combining this with the initialization and read-in techniques,
i.e.\ ESR pulses
to switch the  spin state,
 we have a {\it spin memory} at the ultimate single-spin limit,
 whose relaxation time is just the spin relaxation time. This
relaxation time can be expected to be on the order of $100$'s of
nanoseconds\cite{Kikkawa}, and can be directly measured via the
currents when they switch from high to low due to a spin
flip on the dot\cite{Recher}.

\section{Conclusions}
We have described a scalable scenario for the implementation
 of a solid state quantum computer based on the electron spin
in quantum dots as the qubit. We have shown how electron
spins can be manipulated through their charge (orbital) degrees of
freedom to implement single and two-qubit gates as well as the
possibility of read in/out a single qubit (spin).

\end{article}
\end{document}